# RADIO SCIENCE INVESTIGATIONS FOR THE HEAVY METAL MISSION TO ASTEROID (216) KLEOPATRA

Paolo Tortora,[*] Riccardo Lasagni Manghi,[†] Edoardo Gramigna,[‡] Marco Zannoni,[§] Jan-Erik Wahlund,[**] Jan Bergman,[††]

**This work presents the simulation results of the radio science experiment onboard the proposed Heavy Metal mission to the M-type asteroid (216) Kleopatra. Earth-based radiometric measurements (range and range-rate), complemented by images from the onboard optical camera and by measurements from the intersatellite link between the maincraft and a secondary subcraft, are used in an orbit determination process to assess the attainable accuracy for the mass of Kleopatra and its extended gravity field. Preliminary results indicate that the asteroid mass can be retrieved with a relative accuracy up to $10^{-7}$, while the extended gravity field can be estimated up to degree 10 with a sufficient accuracy to discriminate between internal structure models, satisfying the scientific goals of the mission.**

## INTRODUCTION

Heavy Metal is a proposed ESA mission targeting the main-belt metallic asteroid *(216) Kleopatra* and its two moons. Metallic-rich (M-class) asteroids represent a fairly common type of asteroid, which are mainly located in the middle- and outer-regions of the main belt. So far, very limited observational information is available for such bodies, with only one exploration mission, the NASA Psyche mission[1], set to rendezvous with the largest M-class asteroid *(16) Psyche*.

Belonging to a different dynamic family than Psyche, Kleopatra is the second largest metallic asteroid in the main belt and represents one the most exciting targets for a substantial science mission. Its high bulk density and metallic content, highlighted in multiple studies, provide an extraordinary opportunity to investigate the composition and geology of a possible proto-planetary core or re-accreted asteroid. Its shape also suggests a complex formation mechanism and may shed light on the formation of binary and contact-binary asteroids[2]. Furthermore, the presence of two small moons, Alexhelios and Cleoselene, and the peculiar mass distribution of Kleopatra gives rise to a non-trivial long-term dynamical environment. The structure and composition of the two moons will help determine if they originated from the main body or were captured and provide insight into the formation of asteroid satellites.

[*] Full Professor, Dipartimento di Ingegneria Industriale, Alma Mater Studiorum – Università di Bologna, via Fontanelle 40, 47121, Forlì, Italy.
[†] PostDoc Researcher, Dipartimento di Ingegneria Industriale, Alma Mater Studiorum – Università di Bologna, via Fontanelle 40, 47121, Forlì, Italy.
[‡] PhD Student, Dipartimento di Ingegneria Industriale, Alma Mater Studiorum – Università di Bologna, via Fontanelle 40, 47121, Forlì, Italy.
[§] Assistant Professor, Dipartimento di Ingegneria Industriale, Alma Mater Studiorum – Università di Bologna, via Fontanelle 40, 47121, Forlì, Italy.
[**] Associate Professor, Swedish Institute of Space Physics, Box 537, SE-751 21 Uppsala, Sweden
[††] Senior Scientist, Swedish Institute of Space Physics, Box 537, SE-751 21 Uppsala, Sweden

The study described here focuses on the simulation of the radio science experiment (RSE) onboard the Heavy Metal mission, whose main objective is to estimate the gravity field of the planetary bodies to constrain their interior structure and mass distribution. This estimation is realized by means of an accurate orbit determination process, which makes use of range and range-rate measurements from the tracking link between the Earth ground stations and the Heavy Metal main spacecraft (or maincraft), optical navigation (OPNAV) images collected by the onboard cameras, and the inter-satellite link (ISL) between the maincraft and a secondary smaller spacecraft (or subcraft). Numerical simulations of the orbit determination process for Heavy Metal were performed, providing a preliminary assessment of the achievable accuracy in the estimation of the scientific parameters of interest, by means of a covariance analysis.

**MISSION SCENARIO**

Heavy Metal, consisting of a maincraft and a smaller subcraft, would be launched by an Ariane 6.2 rocket during the launch window of 2036-2037. Using an electric propulsion thruster and a gravity assist maneuver at Mars, the spacecraft are expected to arrive at Kleopatra after a 3- to 5-year cruise. The maincraft is planned to operate near the body for a period of roughly 1 year, after which it may either be sent to the surface for a controlled landing or sent to additional M-class asteroids in an extended mission.

The initial injection orbit of the Heavy Metal maincraft is a prograde circular orbit around Kleopatra (KCO) at a distance of 5000 km and at an inclination of 45° relative to the asteroid equatorial plane. At this distance the spacecraft is already influenced by Kleopatra's extended gravitational field, but the orbit retains its mean inclination, eccentricity, and semi-major axis. During this phase, the orbital elements of the two moons will be characterized in detail.

From its initial orbit, the maincraft will descend down to a 700 km KCO, right outside the orbit of the outer moon Alexhelios, retaining a 45° inclination. The orbit has a precession of -4.9° every orbital period of 72.2 h, which allows to collect detailed surface maps of Kleopatra under various illumination conditions. At this point, the subcraft will separate from the maincraft and insert into a polar orbit through an inclination change maneuver, which will minimize the node drift and facilitate the orbit phasing during fly-bys. The main scientific goal of the subcraft is in fact to complement the measurements collected by the main probe, performing low-altitude fly-bys of the two moons and of Kleopatra, and providing the means for measuring spatial gradients of the gravity, electric, and magnetic fields.

During its mission of roughly three months, the subcraft will progressively descend to lower-altitude and higher-risk orbits, as shown in Figure 1, before impacting on the polar surface of Kleopatra through a controlled crash. In the same time span, the maincraft will support subcraft operations via ISL from higher-altitude orbits. Only once the dynamical environment has been characterized with a sufficient level of detail through close-up observations and gravity measurements, the maincraft will follow the subcraft to lower-altitude orbits, as indicated in Table 1, which summarizes spacecraft operations as a function of the mission phases.

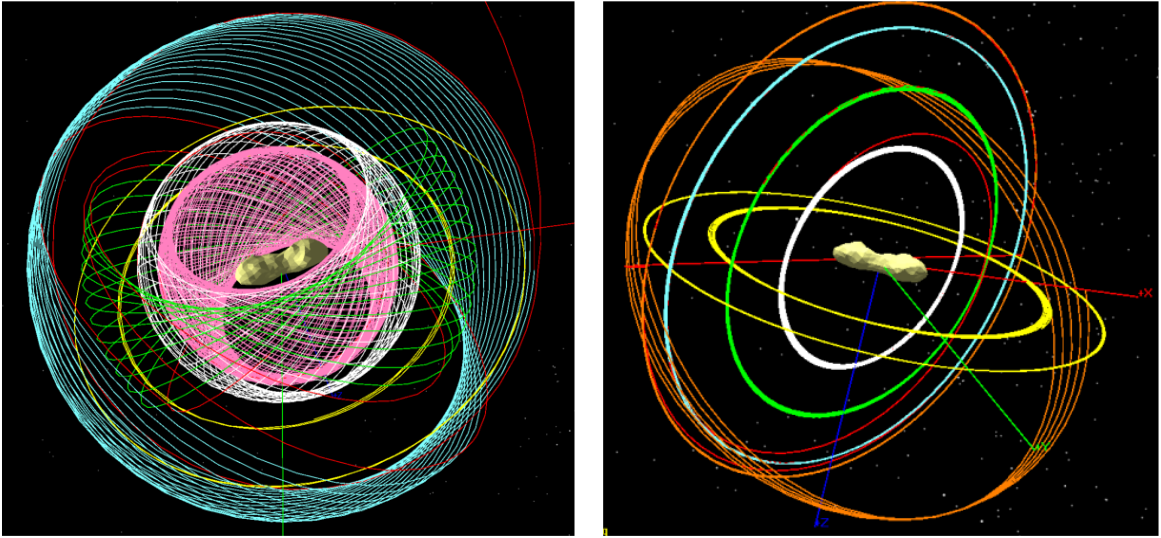

**Figure 1** Orbits of spacecraft and satellites around Kleopatra (Credit OHB). Left: maincraft proximity orbits, corresponding to KCOs with 45° inclination and semi-major axis of 700 km (blue), 550 km (green), 400 km (white), and 260 km (pink); Right: subcraft proximity orbits, corresponding to KCOs with 90° inclination, except for the initial injection orbit, and semi-major axis of 650 km (blue), 499 km (green), and 260 km (white); Transfer orbits are depicted in red, while orbits of the two moons Alexhelios and Cleoselene are shown in yellow.

**Table 1** Reference timeline and simulated orbits for the Heavy Metal mission.

| Phase | Duration | Orbits | Science tasks / comments |
|---|---|---|---|
| A | 2 months | KCO at 5000 km | Intermittent science observations |
| B | 2 months | M: KCO at 700 km<br>S: KCO at 650 km and 499 km | Separation of subcraft (S) and maincraft (M); M: support to subcraft operations via ISL; S: science operations at the moons |
| C | 4 weeks | M: KCO at 550 km<br>S: KCO at 499 km and 260 km | M: science operations at the moons<br>S: science operations at innermost safe orbit |
| D | 5 weeks | M: KCO at 400 km<br>S: elliptic orbits around 260 km | M: support to subcraft operations via ISL<br>S: proximity operations at Kleopatra and controlled crash on the surface |
| E | 4 months | M: KCO at 260 km | M: close operations at target and landing or mission extension |

## SIMULATION PROCEDURE

A radio science experiment represents a particular application of the orbit determination process in which a series of parameters is estimated that completely define the past trajectory of a spacecraft and enable to predict its future evolution. For the Heavy Metal radio science experiment these parameters include, among others, the mass of Kleopatra, its extended gravity field expressed in terms of spherical harmonics, its heliocentric trajectory, and its rotational state.

To assess the formal uncertainty of the estimated parameters before the execution of the mission, numerical simulations are run, which follow the same procedure of the real experiment, using simulated measurements in place of observed data. For this work a "covariance analysis" was performed, meaning that the dynamical and observational models used for generating the simulated observables are the same ones that are used for the estimation process. This approach is of-ten used during mission design phases to get a better understanding of how the spacecraft trajectories and the main design parameters affect the performance of the experiment[3,4].

However, the real uncertainty associated with the estimated parameters is often larger than the formal value estimated within the simulated orbit determination process due to linearization errors, biases in the estimated values induced by unmodelled physical effects, and correlated measurement errors. To account for these factors and obtain realistic expected uncertainties, a series of conservative assumptions were made, including higher noise levels and larger *a priori* uncertainties for the solved-for parameters.

Numerical simulations were performed using NASA/JPL's navigation software, the Mission Analysis, Operations, and Navigation Toolkit Environment (MONTE)[5], whose mathematical formulation is described by Moyer[6], and which is currently (or has been) used for the operations in NASA's deep-space missions, and for radio science experiments data analysis[7,8,9,10].

**DYNAMICAL MODEL**

The heliocentric orbit of Kleopatra and the relative orbits of the Heavy Metal maincraft and subcraft, were integrated numerically using the following gravitational accelerations: relativistic point-mass gravity of the Sun, the Solar System planets and their satellites, the Moon, and Pluto; point-mass gravity of Kleopatra, along with its higher-order gravitational harmonics.

The gravitational coefficients and initial state vectors of the Solar System bodies were taken from JPL's DE440 planetary ephemerides[11], while the initial state of Kleopatra was retrieved from JPL's Small Body database[12] in form of osculating orbital elements.

The rotational model of Kleopatra was described with respect to the ecliptic mean orbit at J2000 (EMO2000) by its pole right ascension α and declination δ coordinates, which are modelled as linear functions of time. Similarly, the prime meridian *w* is modelled as a linear function of time starting from an initial value of zero at the node and having a constant angular frequency.

The gravity field of Kleopatra was expanded up to degree and order 10, which represents the highest observable degree with the currently available shape models[13]. Specifically, the *a* priori values of the Stokes coefficients were derived by Broz[14], which were scaled to a reference radius of 140 km. The GM and Stokes coefficients up to highest degree and order were estimated within the orbit determination filter.

The non-gravitational acceleration due to solar radiation pressure (SRP) was computed using a standard flat-plates model and assuming a simplified spacecraft shape composed by a cuboid bus, deployable solar arrays, and a high-gain antenna (only for the maincraft). Table 2 summarizes the physical properties of the main spacecraft surfaces. SRP modelization errors typically represent one of the main error sources for the non-gravitational accelerations. To mitigate possible mismodelling effects, local SRP scale factors were estimated for each arc as part of the orbit determination process.

**Table 2 Shape models and optical coefficients for the Heavy Metal spacecraft**

| Probe | Component | Area (m$^2$) | Specular | Diffusive |
|---|---|---|---|---|
| Maincraft | High-gain antenna | 3.67 | 0 | 0.327 |
| | Bus (top/side/front) | 2.9 / 4.6 / 4.6 | 0.0735 | 0.2520 |
| | Solar arrays | 43 | 0.038 | 0.052 |
| Subcraft | Bus (top/side/front) | 0.0425 / 0.0875 / 0.0278 | 0.0735 | 0.2520 |
| | Solar arrays | 0.1455 | 0.038 | 0.052 |

**MEASUREMENT MODELS**

The orbit determination process used simulated range and range-rate (Doppler) measurements between the maincraft and the Earth ground stations, assuming a coverage of 8 h/day from a single ESTRACK tracking station. The maincraft will support a multi-frequency tracking link at X- and Ka-bands, which will allow to remove the errors induced by dispersive media such as the solar plasma and the Earth's ionosphere[15,16]. However, we conservatively assumed noise values of 0.07 mm/s for the simulated Doppler measurements at 60 s count time and of 43 cm for the range, which include a safety factor of 2 with respect to the typical accuracy for single-frequency links at X-band[17].

Furthermore, two-way range and Doppler measurements at S-band were simulated for the inter-satellite link between the maincraft and the subcraft. For the ISL measurements we assumed a 20 % duty cycle, equally spread over the subcraft operational life, which corresponds to 1 min of tracking every 5 min of operations. Conservative noise values of 0.05 mm/s and 50 cm were respectively used for the Doppler at 60 s count time and for the range, in agreement with preliminary tests on ISL hardware components for the Hera mission[18].

Optical navigation images were acquired outside of tracking periods, during which the high-gain antenna shall be pointed towards the Earth, at a sampling rate of 1 picture every 5 h. Additionally, only pictures for which the Sun phase angle was below 60° were used within the filter due to operational constraints. The optical observables are represented by the sample and line coordinates of a variable number of surface landmarks that are visible in the picture or alternatively by the coordinates of the Kleopatra centroid when the body size in the image is below 100 pixels. A total of 528 landmarks were generated, equally spaced at latitude and longitude steps of 15° on the surface of Kleopatra. As a reference, the average number of landmarks detected in a single image during the Rosetta mission varied between 50 and 1000 depending on the orbital altitude[19].

The sample and line coordinates of the landmarks (or centroid) were degraded using an additive white gaussian noise with a standard deviation of 2 pixels. Additionally, a camera pointing error was introduced for each picture in the form of three rotations around the camera axes, which were estimated within the orbit determination filter starting from an *a priori* uncertainty of 10 mdeg. Both the pixel noise and the pointing errors represents conservative assumptions with respect to the values reported by other small body missions such as Rosetta[20] or Dawn[21], and account for modelling errors such as landmarks' misdetection, image processing biases, or offsets between the centroid and the target's center of mass.

**FILTER SETUP**

In this study we adopted a multi-arc approach, which is commonly used for the orbit determination of missions towards small bodies[22]. This approach consists in combining the data collected during non-contiguous orbital segments, or "arcs", and analyze them jointly in a weighted least-square batch filter to produce a single set of solved-for parameters. The solution is represented by the estimated values of the parameters and the corresponding covariance matrix, which provides the formal estimation uncertainty. The solved-for parameters, which are summarized in Table 3, can be broadly divided in *global* parameters, which do not vary during the mission and thus affect all tracking arcs, and *local* parameters that only affect the measurements of individual arcs.

**Table 3 Filter setup summary**

| Parameter | Type | *A priori* σ | Comments |
|---|---|---|---|
| *Heavy Metal (maincraft and subcraft)* | | | |
| Position | Local | 10 km | Spacecraft state with respect to Kleopatra. Accounts for ma- |
| Velocity | Local | $10^{-4}$ km/s | neuver errors at the beginning of the arc |
| *Kleopatra* | | | |
| Position | Global | 6.7864 km | From uncertainty of orbital elements[12], scaled by a factor of 5. |
| Velocity | Global | $2.1768 \cdot 10^{-5}$ km/s | |
| GM | Global | $3.1815 \cdot 10^{18}$ kg | From estimated uncertainty[14], scaled by a factor 10 |
| $C_{20}$ | Global | 1.5336 | From power spectrum of estimated values[14], scaled by a factor of 10. Only the degree and order 2 are reported for brevity. |
| $C_{22}$ | Global | $4.4272 \cdot 10^{-1}$ | |
| $S_{22}$ | Global | $4.4272 \cdot 10^{-1}$ | |
| $\alpha_0$ | Global | 5 deg | From estimated uncertainty[14], scaled by a factor of 5. |
| $\delta_0$ | Global | 5 deg | |
| $w_1$ | Global | $1.7241 \cdot 10^{-8}$ deg/s | From rotation period uncertainty[14], scaled by a factor of 5. |
| *SRP* | | | |
| Scale Factor | Local | 1.0 | Uncertainty is 100 % of the acceleration |
| *Pointing error per picture* | | | |
| 3 Rotations | Local | 10 mdeg | Same as Rosetta[20] |
| *Kleopatra landmark positions* | | | |
| Radius | Global | - | 10 % of the average radius |
| Lat. / Long. | Global | 5 deg | |
| Scale factor | Global | 0.1 | 10 % size scale, common to all landmarks |

**RESULTS**

A series of simulations were run to assess the radio science experiment performance as a function of the Heavy Metal mission phase. Specifically, the following scenarios were explored, which assess the information content provided by the individual probes:

1. *Baseline scenario*: only measurements collected by the maincraft are available, namely range and Doppler from the Earth-based tracking link, and OPNAV images;

2. *ISL scenario*: all measurements collected by the maincraft and the subcraft are available, namely range and Doppler from the Earth-based tracking link, OPNAV images, and range and Doppler from the ISL.

Figure 2 shows the estimated uncertainty of Kleopatra's GM for the two simulated scenarios. It can be seen from the left plot that the GM relative accuracy improves by more than two orders of magnitude, from roughly $9 \cdot 10^{-5}$ at the end of phase A to $3 \cdot 10^{-7}$ at the end of phase E. However, the mission requirement of $10^{-7}$ is met only in the *ISL scenario* (right plot), which shows a drastic improvement of the results by the end of phase B thanks to the information content provided by the ISL at the lower-altitude orbits of the subcraft.

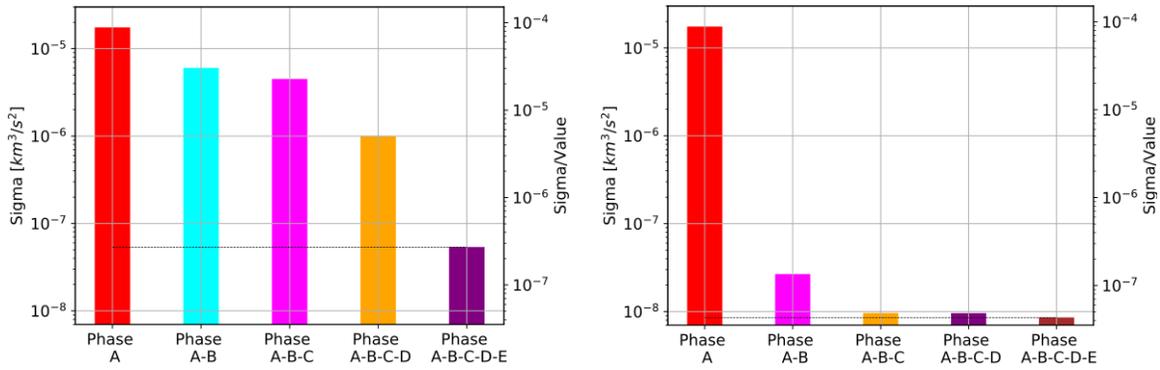

**Figure 2 Estimated uncertainty of Kleopatra's GM. Left:** *baseline scenario;* **right:** *ISL scenario*.

Similar considerations can be drawn for Kleopatra's J$_2$, which is shown in Figure 3. By the end of phase A this parameter can already be estimated with a relative accuracy of 0.07, which improves to values of $2 \cdot 10^{-6}$ and $2 \cdot 10^{-7}$ at the end of mission, respectively for the *baseline* and *ISL* scenarios. For the latter, the largest uncertainty reduction is observed during phases B and C when the ISL range and range-rate measurements are processed.

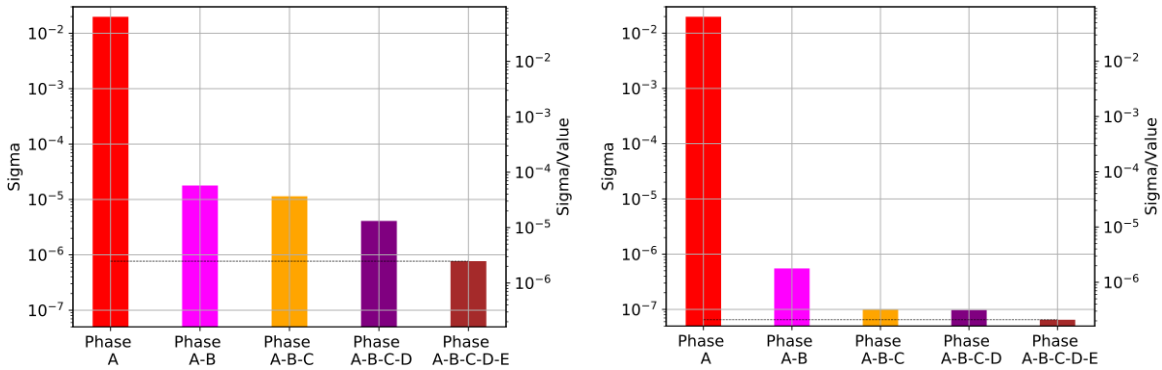

**Figure 3 Estimated uncertainty of Kleopatra's J$_2$. Left:** *baseline scenario;* **right:** *ISL scenario*.

Figure 4 shows Kleopatra's gravity spectra for a spherical harmonic expansion up to degree and order 10. Specifically, the continuous black line represent the root mean square (RMS) value of the Stokes coefficients of all orders for a given degree expansion. Similarly, the colored lines represent the RMS value of the estimated uncertainty of the Stokes coefficients at the end of each mission phase. In the *baseline scenario*, higher-order harmonics can be estimated with a relative accuracy in the order of ~1%, on average, only at the end of the mission. In the *ISL scenario*, comparable accuracies are reached by the end of phase C, while a factor 2 to 5 improvement is observed at the end of mission. This is most likely due to the different observing geometries for the subcraft proximity orbits, which allow to remove most of the ambiguities in the gravity field around Kleopatra's poles.

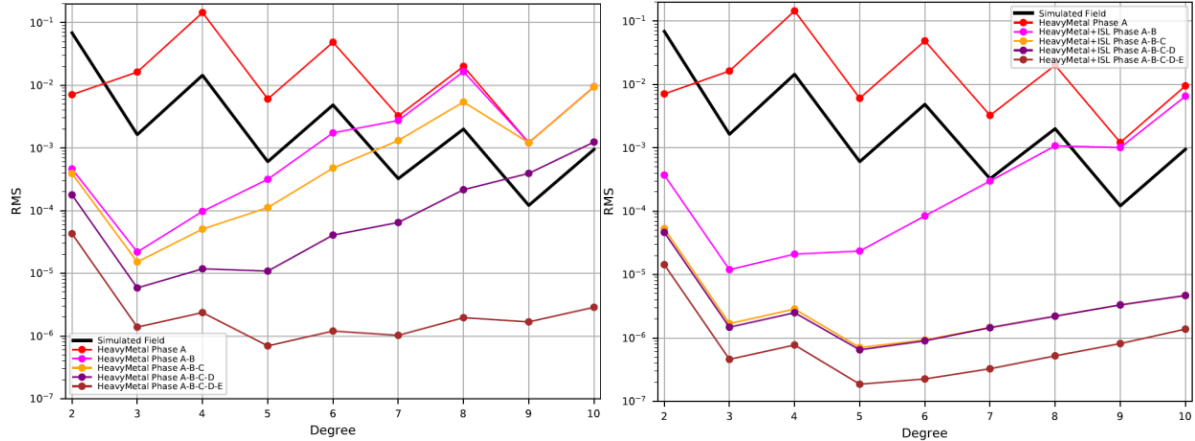

**Figure 4** Kleopatra's gravity spectra. Left: *baseline scenario;* right: *ISL scenario*. Black lines represent the power spectrum of the simulated field (RMS of the Stokes coefficients of all orders for a given degree of spherical harmonics expansion). Colored lines show the estimated formal uncertainties (1-sigma) at the end of phase A (red), phase B (pink), phase C (orange), phase D (violet), and phase E (brown). It should be noted that the subcraft is only active during phases B and C (see Table 1).

The orbit determination simulations also show an excellent sensitivity to the asteroid's pole direction and rotational speed. The former, expressed in terms of right ascension and declination, can be estimated with an accuracy of less than $4.9\times10^{-2}$ deg and $1.5\times10^{-2}$ deg, respectively for the *baseline* and *ISL* scenarios, which correspond to displacements of 51 m and 16 m on the surface of an equivalent-volume sphere. Furthermore, the rotational speed can be estimated with an accuracy in the order of $10^{-11}$ deg/sec, which correspond to roughly 0.05 deg/century.

## CONCLUSIONS

The main result of the presented work is that the radio science experiment for the proposed Heavy Metal mission to the asteroid *(216) Kleopatra* has proved feasible. The results show that, at the end of the nominal mission, the mass of Kleopatra can be estimated with a relative accuracy better than $10^{-7}$. The extended gravity field can be reliably estimated up to degree 10, representing the highest observable degree with the currently available shape models. These high-order gravity terms correspond to density variations with a scale length of 19 km on the surface of Kleopatra. Moreover, the zonal harmonic $J_2$ can be estimated with a relative accuracy better than $10^{-6}$, allowing to discriminate between models of the internal structure with various levels of differentiation, constraining the size and density of a possible internal core. Adding to the main probe a deployable subcraft with ISL capabilities has proven to increase the scientific return of the mission, while also providing critical redundancy during the proximity operations.

Further improvements in the orbit determination accuracy may be obtained exploiting the optical images collected by the onboard cameras of the subcraft, which were not considered for this analysis. Future work may also focus on improving the fidelity of the simulations by including the two moons Alexhelios and Cleoselene within the dynamical model and estimating their GMs and ephemerides within the orbit determination process through dedicated low-altitude flybys.

## ACKNOWLEDGEMENTS

EG, RLM, MZ, and PT wish to acknowledge *Caltech* and the *NASA Jet Propulsion Laboratory* for granting the University of Bologna a license to an executable version of MONTE Project Edition S/W.

[22] Lasagni Manghi, R., Zannoni, M., Tortora, P. & Modenini, D., 2019. Measuring The Mass of A Main Belt Comet: Proteus Mission. Turin, IEEE 5th International Workshop on Metrology for AeroSpace (MetroAeroSpace).